\documentclass[12pt]{article}
\usepackage{amsmath}
\usepackage{amssymb}
\usepackage{geometry}
\usepackage{mathrsfs}

\newtheorem{theorem}{Theorem}

\newtheorem{remark}[theorem]{Remark}

\input{tcilatex}
\begin{document}

\title{\textbf{The isotropic lines of }$%
\mathbb{Z}
_{d}^{2}$}
\author{Olivier Albouy$^{1,2,3}$}
\date{}
\maketitle

$^{1}$ Universit\'{e} de Lyon, F-69622, Lyon, France

$^{2}$ Universit\'{e} Lyon 1, Villeurbanne, France

$^{3}$ CNRS/IN2P3, UMR5822, Institut de Physique Nucl\'{e}aire de Lyon,
France

\bigskip

E-mail: o.albouy@ipnl.in2p3.fr

\bigskip

\begin{center}
\textbf{Abstract}
\end{center}

We show that the isotropic lines in the lattice $%
\mathbb{Z}
_{d}^{2}$ are the Lagrangian submodules of that lattice and we give their
number together with the number of them through a given point of the
lattice. The set of isotropic lines decompose into orbits under the action
of $\func{SL}(2,%
\mathbb{Z}
_{d})$. We give an explicit description of those orbits as well as their
number and their respective cardinalities. We also develop two group actions
on the group $\Sigma _{\mathscr{D}}(M)$ related to the topic.

\bigskip



\bigskip

\section*{Introduction}

Wigner distributions are a major tool of quantum mechanics. They offer a
useful, alternative way besides density matrices of representing pure and
mixed states of a quantum system. But whereas in the continuous phase space
those distributions are well-defined \cite{Wigner.1932}\cite{Groot.1975},
there is still a need for a sound mathematical definition over a discrete
phase space.\ In particular, the structure of such a phase space is of some
importance. In 1974, Buot introduced a Wigner distribution over an $d\times d
$ phase space with $d$ an odd integer \cite{Buot.1974}. In 1980, Hannay and
Berry followed another approach to build a Wigner distribution over a $%
2d\times 2d$ lattice \cite{Hann.Berr.1980}.\ Still in another way, in 2004,
Gibbons \textit{et al.} constructed Wigner distributions over a finite field
parametrised lattice \cite{Gibbons.al.2004}.

More recently, in their way to set up discrete Wigner distributions on the
discrete phase space $%
\mathbb{Z}
_{d}^{2}$, with $%
\mathbb{Z}
_{d}$ the set of integers modulo $d$, Chaturvedi \textit{et al} \cite%
{Chaturvedi.al.06} encountered undetermined signs $S(q,p)$, one at each
point $(q,p)$\ of the lattice. A natural question then arises: To what
extent can these signs be fixed by demanding that averages of Wigner
distributions over isotropic lines in the lattice yield probabilities, where
an isotropic line is a set of $d$ points on the lattice such that the
symplectic product of any two of them is $0$ (modulo $d$). In order to
answer this and related questions one needs a detailed knowledge of the
structure of the isotropic lines in $%
\mathbb{Z}
_{d}^{2}$. In particular, it would be useful to know their number as a whole
or with special conditions and also how they arrange in orbits under the
action of the symplectic group $\func{SL}(2,%
\mathbb{Z}
_{d})$.

This communication is only concerned with the mathematical properties of the
isotropic lines in $%
\mathbb{Z}
_{d}^{2}$. In Section~\ref{Section: Number of IL}, we derive the number of
isotropic lines in $%
\mathbb{Z}
_{d}^{2}$ and then in Section~\ref{Section: Number of IL thru x}\ the number
of them through a given point of the lattice. This should be compared with
the results obtained by Havlicek and Saniga in \cite{Hav.San.07} and \cite%
{Hav.San.08} about the number of projective points in the lattice and the
number of them under the same condition. In Section~\ref{Section: Orbits},
we give a full description of the orbits of isotropic lines under the action
of $\func{SL}(2,%
\mathbb{Z}
_{d})$ with the help of some parameters. All that is achieved on the basis
of a work by the author on symplectic reduction of matrices and Lagrangian
submodules \cite{Albouy.08}. In a fourth section, we develop two group
actions on the group $\Sigma _{\mathscr{D}}(M)$ relevant to the
understanding of that latter group. For any result appearing in this
communication without proof, the reader is referred to \cite{Albouy.08}.

To end this introduction, we note two features of the results presented
here. On the one hand, they do not depend on the parity of $d$, contrary to
what happened in \cite{Buot.1974}\ and \cite{Hann.Berr.1980}. On the other
hand, they have direct relevance to the commuting subgroups of the Pauli
group for a general $d$. That latter feature has been of great importance in
the building up of both Wigner ditrbutions and mutually unbiased bases \cite%
{Gibbons.al.2004}\cite{Bandyopadhyay.al.2002}.

\section{The number of isotropic lines\label{Section: Number of IL}}

Let $\omega $ denote the symplectic product of two vectors of $%
\mathbb{Z}
_{d}^{2}$. With matrices, it consists in computing a determinant:%
\begin{equation}
\omega ((\alpha ,\beta ),(\gamma ,\delta ))=\left\vert 
\begin{array}{cc}
\alpha  & \gamma  \\ 
\beta  & \delta 
\end{array}%
\right\vert =\alpha \delta -\beta \gamma .
\end{equation}%
The orthogonal of a submodule $M$ of $%
\mathbb{Z}
_{d}^{2}$ will be denoted $M^{\omega }$:%
\begin{equation}
M^{\omega }=\{x\in 
\mathbb{Z}
_{d}^{2};\forall y\in M,\omega (x,y)=0\}.
\end{equation}%
Isotropic submodules are defined to satisfy the set inclusion $M\subset
M^{\omega }$. Lagrangian submodules are the maximal isotropic submodules for
inclusion, what is equivalent to $M=M^{\omega }$.

In a first time, we are going to find the number of isotropic lines in $%
\mathbb{Z}
_{d}^{2}$ for $d$ a power of a prime, say $d=p^{s}$, $s\geq 1$. The way we
derive this number is a strict application of Theorem 7\ in \cite{Albouy.08}%
. This brings about a hint for Section \ref{Section: Actions}, but we shall
also see that there exists a shortcut. We then address the case of a general 
$d$.

\subsection{Special case: $d$ a power of a prime}

Let $\widetilde{s}=\left\lfloor s/2\right\rfloor $, the floor part of $s/2$.
As shown in \cite{Albouy.08}, for any Lagrangian submodule $M$, there exist $%
S\in \func{SL}(2,%
\mathbb{Z}
_{p^{s}})$ and $k\in \{0,\ldots ,\widetilde{s}\}$ such that $M$ is linearly
generated by the column vectors of%
\begin{equation}
S\times \left( 
\begin{array}{cc}
p^{k} & 0 \\ 
0 & p^{s-k}%
\end{array}%
\right) .  \label{Lagr sm decomposition}
\end{equation}%
In other words, with $S_{1}$ and $S_{2}$ the two column vectors of $S$, $%
(S_{1},S_{2})$ is a symplectic computational basis of $(%
\mathbb{Z}
_{p^{s}})^{2}$ and $M$ is the set of all linear combinations of $p^{k}S_{1}$
and $p^{s-k}S_{2}$ with coefficients in $%
\mathbb{Z}
_{p^{s}}$. As a converse, any submodule thus generated is Lagrangian. In
fact, the number $k$ is a property of $M$, that is to say for any convenient
pair $(S,k^{\prime })$ in order to generate $M$ as in (\ref{Lagr sm
decomposition}), we have $k^{\prime }=k$. We will denote $\mathbf{O}%
_{k}(p^{s})$ the set of all Lagrangian submodules thus obtained for a given $%
k$ and $S$ varying. The cardinality of any $M\in \mathbf{O}_{k}(p^{s})$ is%
\begin{equation}
p^{(s-1)-(k-1)}p^{(s-1)-(s-k-1)}=p^{s}.
\end{equation}%
Let $\ell $ be an isotropic line and $\left\langle \ell \right\rangle $ the
submodule it generates, the set of all finite linear combinations of vectors
of $\ell $. Any two vectors in $\left\langle \ell \right\rangle $ are
orthogonal and hence $\left\langle \ell \right\rangle $ is an isotropic
submodule containing at least $p^{s}$ vectors. Thus isotropic lines and
Lagrangian submodules are the same.

The number of free vectors $x$ in $%
\mathbb{Z}
_{p^{s}}$ is $p^{2s}-p^{2(s-1)}$. The number of vectors $y$ such that for a
given free $x$ we have $\omega (x,y)=1$ is $p^{s}$. The number of pairs $%
(x,y)$ such that $\omega (x,y)=1$ is the product of the two previous ones:%
\begin{equation}
n_{\omega }=\left\vert \func{SL}(2,%
\mathbb{Z}
_{p^{s}})\right\vert =p^{3s}-p^{3s-2}.
\end{equation}%
Several symplectic matrices $S$ may give rise to the same submodule in $%
\mathbf{O}_{k}(p^{s})$ according to the form (\ref{Lagr sm decomposition}).
Let $k\in \{0,\ldots ,\widetilde{s}\}$ and $M\in \mathbf{O}_{k}(p^{s})$. Let 
$\Sigma _{\mathscr{D}}(M)$ be the matrix group of the changes of
computational basis such that if $P\in \Sigma _{\mathscr{D}}(M)$ and if $M$
is generated by the column vectors of the matrix given in (\ref{Lagr sm
decomposition}), then $M$ is also generated by the column vectors of the
matrix%
\begin{equation}
SP\times \left( 
\begin{array}{cc}
p^{k} & 0 \\ 
0 & p^{s-k}%
\end{array}%
\right) ,
\end{equation}%
where $SP$ need not be symplectic. In fact, we derived in \cite{Albouy.08}\
that the group $\Sigma _{\mathscr{D}}(M)$ is completely determined by the
value of $k$. So, the number of symplectic matrices that give rise to a
given $M\in \mathbf{O}_{k}(p^{s})$ is%
\begin{equation}
n_{\mathscr{D}}(k)=\left\vert \Sigma _{\mathscr{D}}(M)\cap \func{SL}(2,%
\mathbb{Z}
_{p^{s}})\right\vert 
\end{equation}%
and hence%
\begin{equation}
\left\vert \mathbf{O}_{k}(p^{s})\right\vert =\frac{n_{\omega }}{n_{%
\mathscr{D}}(k)}.  \label{Cardinality of O_k medium}
\end{equation}%
Let us suppose that $2k<s$. In $\Sigma _{\mathscr{D}}(M)$, the number of
matrices with determinant $1$ is the same as the number of matrices with any
other (invertible) determinant. Indeed, if $u\in U(%
\mathbb{Z}
_{p^{s}})$ and $P=(P_{1}|P_{2})\in \Sigma _{\mathscr{D}}(M)\cap \func{SL}(2,%
\mathbb{Z}
_{p^{s}})$, with $P_{1}$ and $P_{2}$ the first and second columns of $P$
respectively, then $(uP_{1}|P_{2})\in \Sigma _{\mathscr{D}}(M)$ but with
determinant $u$. This transformation is injective so that the number of
matrices in $\Sigma _{\mathscr{D}}(M)$ with determinant $u$ is greater than
or equal to the number of matrices in $\Sigma _{\mathscr{D}}(M)$ with
determinant $1$. The converse inequality may be shown the same way. So we
have 
\begin{subequations}
\label{nD_k}
\begin{eqnarray}
n_{\mathscr{D}}(k) &=&\frac{\left\vert \Sigma _{\mathscr{D}}(M)\right\vert }{%
\left\vert U(%
\mathbb{Z}
_{p^{s}})\right\vert }=\frac{(p^{s}-p^{s-1})^{2}\cdot
p^{(s-1)-(s-2k-1)}\cdot p^{s}}{p^{s}-p^{s-1}}=(p^{s}-p^{s-1})p^{s+2k}
\label{nD_k 1} \\
&=&p^{2s}(p^{2k}-p^{2k-1})
\end{eqnarray}%
and so 
\end{subequations}
\begin{equation}
\left\vert \mathbf{O}_{k}(p^{s})\right\vert =\frac{p^{s}-p^{s-2}}{%
p^{2k}-p^{2k-1}}=p^{s-2k-1}(p+1).  \label{Cardinality of O_k}
\end{equation}%
If $2k=s$ (what supposes that $s$ is even), then $\Sigma _{\mathscr{D}%
}(M)\cap \func{SL}(2,%
\mathbb{Z}
_{p^{s}})=\func{SL}(2,%
\mathbb{Z}
_{p^{s}})$ and so%
\begin{equation}
\left\vert \mathbf{O}_{s/2}(p^{s})\right\vert =\frac{n_{\omega }}{n_{\omega }%
}=1.  \label{Cardinality of O_s/2}
\end{equation}

For $s$ odd, then $2\widetilde{s}=s-1$,%
\begin{equation}
\sum_{k=0}^{\widetilde{s}}p^{-2k}=\frac{1-p^{-2(\widetilde{s}+1)}}{1-p^{-2}}=%
\frac{p^{2(\widetilde{s}+1)}-1}{p^{2(\widetilde{s}+1)}-p^{2\widetilde{s}}}=%
\frac{p^{s+1}-1}{p^{s+1}-p^{s-1}}
\end{equation}%
and hence the number of isotropic lines is%
\begin{equation}
n_{L}(p^{s})=\sum_{k=0}^{\widetilde{s}}\left\vert \mathbf{O}%
_{k}(p^{s})\right\vert =p^{s-1}(p+1)\frac{p^{s+1}-1}{p^{s+1}-p^{s-1}}=\frac{%
p^{s+1}-1}{p-1}.
\end{equation}%
If $s$ is even, then $2\widetilde{s}=s$,%
\begin{equation}
\sum_{k=0}^{\widetilde{s}-1}p^{-2k}=\frac{1-p^{-2\widetilde{s}}}{1-p^{-2}}=%
\frac{p^{2\widetilde{s}}-1}{p^{2\widetilde{s}}-p^{2\widetilde{s}-2}}=\frac{%
p^{s}-1}{p^{s}-p^{s-2}}
\end{equation}%
and hence the number of isotropic lines is again 
\begin{multline}
n_{L}(p^{s})=\sum_{k=0}^{\widetilde{s}-1}\left\vert \mathbf{O}%
_{k}(p^{s})\right\vert +1=p^{s-1}(p+1)\frac{p^{s}-1}{p^{s}-p^{s-2}}+1 \\
=p\frac{p^{s}-1}{p-1}+1=\frac{p^{s+1}-p+p-1}{p-1}=\frac{p^{s+1}-1}{p-1}.
\end{multline}

\subsection{General case: $d$ any integer $\geq 2$}

Now let $d$ be any integer greater than or equal to $2$ and%
\begin{equation}
d=\prod_{i\in I}p_{i}^{s_{i}}  \label{PFD of N}
\end{equation}%
be the prime factor decomposition of $d$. Due to the Chinese remainder
theorem, we can study the structure of an isotropic line $\ell $ in each of
the Chinese factor $(%
\mathbb{Z}
_{p_{i}^{s_{i}}})^{2}$. For every $i\in I$, let $\ell _{i}=\pi _{p_{i}}(\ell
)$ be the $i$-th Chinese projection of $\ell $. As a subgroup of $(%
\mathbb{Z}
_{p_{i}^{s_{i}}})^{2}$, $\left\langle \ell _{i}\right\rangle $ has
cardinality a power of $p_{i}$, say $p_{i}^{t_{i}}$. As an isotropic
submodule of $(%
\mathbb{Z}
_{p_{i}^{s_{i}}})^{2}$, $\left\langle \ell _{i}\right\rangle $ is included
in a Lagrangian submodule and then $t_{i}\leq s_{i}$. So%
\begin{equation}
d=\left\vert \ell \right\vert \leq \prod_{i\in I}\left\vert \ell
_{i}\right\vert \leq \prod_{i\in I}p_{i}^{t_{i}}\leq d,
\end{equation}%
what proves that $t_{i}=s_{i}$. Moreover, if $\ell _{i}\subsetneq
\left\langle \ell _{i}\right\rangle $ for some $i$, the second inequality
just above would be strict, what is impossible and so $\ell
_{i}=\left\langle \ell _{i}\right\rangle $ is a Lagrangian submodule of $(%
\mathbb{Z}
_{p_{i}^{s_{i}}})^{2}$. As to the converse, for all $i\in I$, let $\ell
_{i}^{\prime }$ be a Lagrangian submodule of $(%
\mathbb{Z}
_{p_{i}^{s_{i}}})^{2}$. The set $\ell ^{\prime }$ of all vectors $x\in 
\mathbb{Z}
_{d}^{2}$ such that for all $i$, $\pi _{p_{i}}(x)\in \ell _{i}^{\prime }$,
is an isotropic set with cardinality $d$, namely an isotropic line. The
reader may check that the maps $\ell \mapsto (\ell _{i})_{i\in I}$ and $%
(\ell _{i}^{\prime })_{i\in I}\mapsto \ell ^{\prime }$ thus defined are
reciprocal of one another.

So, isotropic lines and Lagrangian submodules are the same sets of $%
\mathbb{Z}
_{d}^{2}$ and the number of isotropic lines of $%
\mathbb{Z}
_{d}^{2}$ is%
\begin{equation}
n_{L}(d)=\prod_{i\in I}n_{L}\left( p_{i}^{s_{i}}\right) =\prod_{i\in I}\frac{%
p_{i}^{s_{i}+1}-1}{p_{i}-1}.  \label{Number of IL for any N}
\end{equation}

\begin{remark}
In (\ref{Lagr sm decomposition}), the left-hand-side factor was a symplectic
matrix. But in fact, any invertible matrix would be convenient since we are
to consider all the linear combinations of the columns in the product. Thus
we could have calculated the cardinality of an orbit as%
\begin{equation}
\frac{n_{\omega }\left\vert U(%
\mathbb{Z}
_{p^{s}})\right\vert }{\left\vert \Sigma _{\mathscr{D}}(M)\right\vert }\text{
instead of }\frac{n_{\omega }}{\left( \left\vert \Sigma _{\mathscr{D}%
}(M)\right\vert /\left\vert U(%
\mathbb{Z}
_{p^{s}})\right\vert \right) }
\end{equation}%
and the argument between (\ref{Cardinality of O_k medium}) and (\ref{nD_k})
could have been avoided.
\end{remark}

\begin{remark}
Let us assume that $s$\ is even. It should be noticed that the formula for
the cardinality of $\mathbf{O}_{k}(p^{s})$ given in (\ref{Cardinality of O_k}%
) is not valid for $k=s/2$. Indeed, equation (\ref{Cardinality of O_k})
gives $1+1/p$ for that particular value of $k$, what is even not an integer.
Equivalently, $n_{\mathscr{D}}(k)$ and $\left\vert \Sigma _{\mathscr{D}%
}(M)\right\vert $ have no unique expression for all values of $k$. This must
be traced back to the behaviour of $\Sigma _{\mathscr{D}}(M)$ when $k$ is
ranging up to $s/2$ (see~\cite{Albouy.08}).
\end{remark}

\section{The number of lines through a given point\label{Section: Number of
IL thru x}}

We now give the number of isotropic lines through a given point of the
lattice. We suppose that $d=p^{s}$ is a power of a prime. Let $x\in 
\mathbb{Z}
_{d}^{2}$ and let $t=v_{p}(x)$ be the $p$-valuation of $x$. Since all the
vectors in an isotropic line $\ell \in \mathbf{O}_{k}(p^{s})$ have $p$%
-valuation at least $k$, the vector $x$ cannot be in $\ell $ unless $k\leq t$%
. Let us assume that $k$ is such that $s-k\leq t$, what implies that $k\leq
t $. Then for any computational basis $(f_{1},f_{2})$, symplectic or not, $x$
is a linear combination of $p^{k}f_{1}$ and $p^{s-k}f_{2}$. Hence%
\begin{equation}
\forall k\in \{0,\ldots ,\left\lfloor s/2\right\rfloor \},\forall \ell \in 
\mathbf{O}_{k}(p^{s}),(k\geq s-t\Longrightarrow x\in \ell ).
\end{equation}%
That case can occur only if $t\geq \left\lceil s/2\right\rceil $, the
ceiling part of $s/2$. Now, let us assume that $k$ is such that $k\leq t<s-k$%
. Thus $2k<s$ and we search for the symplectic computational bases $%
(f_{1},f_{2})$ such that $x$ is a linear combination of $p^{k}f_{1}$ and $%
p^{s-k}f_{2}$. Let $(f_{1},f_{2})$ be a symplectic computational basis and $%
x=af_{1}+bf_{2}$. Since%
\begin{equation}
v_{p}(\omega (x,f_{2}))=v_{p}(a)\geq t\geq k,
\end{equation}%
we have no extra conditions on the choice of $f_{2}$. But we must have%
\begin{equation}
v_{p}(\omega (x,f_{1}))=v_{p}(b)\geq s-k,
\end{equation}%
what shows that in a symplectic basis where $x=(p^{t},0)$, $f_{1}$ must be
of the form%
\begin{equation}
f_{1}=(\alpha ,\beta p^{s-k-t}),
\end{equation}%
with $\alpha ,\beta \in 
\mathbb{Z}
_{d}$. The number of suitable vectors $f_{1}$ is%
\begin{equation}
(p^{s}-p^{s-1})\cdot p^{(s-1)-(s-k-t-1)}=(p^{s}-p^{s-1})p^{k+t}.
\end{equation}%
The number of suitable vectors $f_{2}$ for a given $f_{1}$ is $p^{s}$. Then
the number of suitable, symplectic computational bases $(f_{1},f_{2})$ is $%
(p^{s}-p^{s-1})p^{s+k+t}$. Moreover, if $f$ is a convenient basis and%
\begin{equation}
\left\langle p^{k}f_{1},p^{s-k}f_{2}\right\rangle =\left\langle
p^{k}f_{1}^{\prime },p^{s-k}f_{2}^{\prime }\right\rangle ,
\end{equation}%
then $f^{\prime }$ is convenient too. With (\ref{nD_k 1}), we deduce that
the number of isotropic lines in $\mathbf{O}_{k}(p^{s})$ containing $x$ is%
\begin{equation}
\frac{(p^{s}-p^{s-1})p^{s+k+t}}{(p^{s}-p^{s-1})p^{s+2k}}=p^{t-k}.
\end{equation}%
Thus, if $t<\left\lceil s/2\right\rceil $, the number of isotropic lines
containing $x$ is%
\begin{equation}
\sum_{k=0}^{t}p^{t-k}=p^{t}\cdot \frac{1-p^{-(t+1)}}{1-p^{-1}}=\frac{%
p^{t+1}-1}{p-1}.
\end{equation}%
If $t\geq \left\lceil s/2\right\rceil $ and $\widetilde{s}=\left\lfloor
s/2\right\rfloor $, the number of isotropic lines containing $x$ is%
\begin{equation}
\sum_{k=0}^{s-t-1}p^{t-k}+\sum_{k=s-t}^{\widetilde{s}}\left\vert \mathbf{O}%
_{k}(p^{s})\right\vert .  \label{Number isolines x}
\end{equation}%
The first term is equal to%
\begin{equation}
p^{t}\cdot \frac{1-p^{-(s-t)}}{1-p^{-1}}=\frac{p^{t+1}-p^{2t-s+1}}{p-1}.
\end{equation}%
For $s$ odd, then $2\widetilde{s}=s-1$,%
\begin{equation}
\sum_{k=s-t}^{\widetilde{s}}p^{-2k}=p^{-2(s-t)}\cdot \frac{1-p^{-2(%
\widetilde{s}-s+t+1)}}{1-p^{-2}}=\frac{p^{2t-s+1}-1}{p^{s-1}(p^{2}-1)},
\end{equation}%
and the second term in (\ref{Number isolines x}) is equal to%
\begin{equation}
p^{s-1}(p+1)\frac{p^{2t-s-1}-1}{p^{s-1}(p^{2}-1)}=\frac{p^{2t-s+1}-1}{p-1}.
\end{equation}%
For $s$ even, then $2\widetilde{s}=s$,%
\begin{equation}
\sum_{k=s-t}^{\widetilde{s}-1}p^{-2k}=p^{-2(s-t)}\cdot \frac{1-p^{-2(%
\widetilde{s}-1-s+t+1)}}{1-p^{-2}}=\frac{p^{2t-s+1}-p}{p^{s-1}(p^{2}-1)},
\end{equation}%
and the second term in (\ref{Number isolines x}) is again%
\begin{equation}
p^{s-1}(p+1)\frac{p^{2t-s+1}-p}{p^{s-1}(p^{2}-1)}+1=\frac{p^{2t-s+1}-1}{p-1}.
\end{equation}%
Hence, in any case, the number of isotropic lines containing some given
vector $x$ with $p$-valuation $t$ is%
\begin{equation}
n_{L}(p^{s};x)=n_{L}(p^{s};t)=\frac{p^{t+1}-1}{p-1}.
\end{equation}%
In particular,%
\begin{equation}
n_{L}(p^{s};t=0)=1\text{ and }n_{L}(p^{s};t=s)=n_{L}(p^{s}).
\end{equation}%
That is to say the sole isotropic line containing a free vector is the
submodule it generates and every isotropic line goes through the null vector.

If $d$ is not necessarily a power of a prime, then with (\ref{PFD of N}) and
for all $i$, $t_{i}=v_{p_{i}}(x)$, we obtain that the number of isotropic
lines containing $x$ is%
\begin{equation}
n_{L}(d;x)=n_{L}(d;(t_{i})_{i\in I})=\prod_{i\in I}\frac{p_{i}^{t_{i}+1}-1}{%
p_{i}-1}.
\end{equation}

\section{Orbits under the action of $\func{SL}(2,%
\mathbb{Z}
_{d})$\label{Section: Orbits}}

As in Section \ref{Section: Number of IL}, we first suppose that $d$ is a
power of a prime, say $d=p^{s}$, $s\geq 1$. Then it is obvious from (\ref%
{Lagr sm decomposition}) that the orbits of the left-action of $\func{SL}(2,%
\mathbb{Z}
_{p^{s}})$ among the isotropic lines are the $\mathbf{O}_{k}(p^{s})$. Their
number is $\left\lfloor s/2\right\rfloor +1$ and we have already seen what
their cardinalities are in (\ref{Cardinality of O_k}) and (\ref{Cardinality
of O_s/2}).

Now if $d$ is a composite integer as in (\ref{PFD of N}), then the set of
the orbits is parametrised by%
\begin{equation}
k=(k_{i})_{i\in I}\in \prod_{i\in I}\{0,\ldots ,\left\lfloor
s_{i}/2\right\rfloor \}
\end{equation}%
and the orbit with index $k$ is%
\begin{equation}
\mathbf{O}_{k}(d)=\left\{ \ell \subset 
\mathbb{Z}
_{d}^{2};\left\vert \ell \right\vert =N,\pi _{p_{i}}(\ell )\in \mathbf{O}%
_{k_{i}}(p_{i}^{s_{i}})\right\} .
\end{equation}%
The number of orbits is%
\begin{equation}
\prod_{i\in I}\left( \left\lfloor s_{i}/2\right\rfloor +1\right) ,
\end{equation}%
and the cardinality of one of them is%
\begin{equation}
\left\vert \mathbf{O}_{k}(d)\right\vert =\prod_{i\in I}\left\vert \mathbf{O}%
_{k_{i}}(p_{i}^{s_{i}})\right\vert .
\end{equation}%
\textbf{Example}\quad \textit{Let us suppose that }$d$\textit{\ contains no
square factor, that is to say in (\ref{PFD of N}), for all }$i\in I$\textit{%
, }$s_{i}=1$\textit{. According to (\ref{Lagr sm decomposition}), with }$k$%
\textit{\ necessarily equal to }$0$\textit{, the isotropic lines are the
submodules that can be generated by a single free vector. These submodules
are called the projective points of }$%
\mathbb{Z}
_{d}^{2}$\textit{. With (\ref{Number of IL for any N}), we find that the
number of isotropic lines is}%
\begin{equation}
n_{L}(d)=\prod_{i\in I}(p_{i}+1).
\end{equation}%
\textit{They all belong to the sole orbit under the action of }$\func{SL}(2,%
\mathbb{Z}
_{d})$\textit{\ corresponding to }$k_{i}=0$\textit{\ for all }$i$\textit{.}

\section{Some group actions on $\Sigma _{\mathscr{D}}(M)$\label{Section:
Actions}}

In this section, we assume that $d=p^{s}$ is a power of a prime. In order to
establish equation (\ref{nD_k}), we showed that the number of matrices in $%
\Sigma _{\mathscr{D}}(M)$ with determinant $1$ is the same as the number of
matrices in the same set with any other (invertible) determinant. The simple
reasoning we used was enough in the frame of Section~\ref{Section: Number of
IL}. But we are going to introduce here two other group actions that are
linked to that point and to Remarks 1 and 2. Let $\rho _{0}$ be the action
of $U(%
\mathbb{Z}
_{p^{s}})$ on $\Sigma _{\mathscr{D}}(M)$ defined by%
\begin{equation}
\forall u\in U(%
\mathbb{Z}
_{p^{s}}),\forall P=(P_{1}|P_{2})\in \Sigma _{\mathscr{D}}(M),\rho
_{0}(u)\cdot P=(uP_{1}|u^{-1}P_{2})
\end{equation}%
and $\rho _{1}$ the action of $U(%
\mathbb{Z}
_{p^{s}})^{2}$ on $\Sigma _{\mathscr{D}}(M)$ defined by%
\begin{equation}
\forall (u_{1},u_{2})\in U(%
\mathbb{Z}
_{p^{s}})^{2},\forall P=(P_{1}|P_{2})\in \Sigma _{\mathscr{D}}(M),\rho
_{1}(u_{1},u_{2})\cdot P=(u_{1}P_{1}|u_{2}P_{2}).
\end{equation}%
All the orbits of $\rho _{0}$ (resp.~$\rho _{1}$) have the same cardinality,
namely $\left\vert U(%
\mathbb{Z}
_{p^{s}})\right\vert =p^{s}-p^{s-1}$ (resp.~$\left\vert U(%
\mathbb{Z}
_{p^{s}})\right\vert ^{2}$). In a given orbit of $\rho _{0}$, every matrix
has the same determinant. Since $%
\mathbb{Z}
_{p^{s}}$ is a commutative ring, those two actions "commute":%
\begin{equation}
\rho _{1}(u_{1},u_{2})\cdot (\rho _{0}(u)\cdot P)=\rho _{0}(u)\cdot (\rho
_{1}(u_{1},u_{2})\cdot P).
\end{equation}

Let $(u_{1},u_{2}),(v_{1},v_{2})\in U(%
\mathbb{Z}
_{p^{s}})^{2}$ such that $u_{1}u_{2}=v_{1}v_{2}$, that is to say%
\begin{equation}
\forall P\in \Sigma _{\mathscr{D}}(M),\det (\rho _{1}(u_{1},u_{2})\cdot
P)=\det (\rho _{1}(v_{1},v_{2})\cdot P).
\end{equation}%
With $\lambda =u_{2}v_{2}^{-1}=u_{1}^{-1}v_{1}\in U(%
\mathbb{Z}
_{p^{s}})$, we have%
\begin{equation}
(v_{1},v_{2})=(\lambda u_{1},\lambda ^{-1}u_{2}).
\end{equation}%
Thus we have a kind of a discrete Hopf fibration. It is given by the action $%
h$ of $U(%
\mathbb{Z}
_{p^{s}})$ on $U(%
\mathbb{Z}
_{p^{s}})^{2}$ defined by%
\begin{equation}
\forall \lambda \in U(%
\mathbb{Z}
_{p^{s}}),\forall (u_{1},u_{2})\in U(%
\mathbb{Z}
_{p^{s}})^{2},h(\lambda )\cdot (u_{1},u_{2})=(\lambda u_{1},\lambda
^{-1}u_{2}).  \label{Hopf fibration}
\end{equation}%
Moreover, the action $\rho =\rho _{1}/(h,\rho _{0})$ of $U(%
\mathbb{Z}
_{p^{s}})^{2}/h$ on $\Sigma _{\mathscr{D}}(M)/\rho _{0}$ is well-defined.
For any $u\in U(%
\mathbb{Z}
_{p^{s}})$, let%
\begin{equation}
D_{u}=\left\{ P\in \Sigma _{\mathscr{D}}(M);\det P=u\right\} .
\end{equation}%
Every orbit of $\rho $ is transversal to $D_{u}/\rho _{0}$. Indeed, let $P$
be in some orbit $O$\ of $\rho _{1}$ with some determinant $v$. Then $%
(uv^{-1}P_{1}|P_{2})$ is in $O$ with determinant $u$ so that there is at
least one orbit of $\rho _{0}$ in $D_{u}\cap O$. Then if $P$ and $%
Q=(u_{1}P_{1}|u_{2}P_{2})$ are in $O$ and have the same determinant, then $%
u_{2}=u_{1}^{-1}$ and thus $P$ and $Q$ are in the same orbit of $\rho _{0}$.

As a conclusion, we have a partition $E=\left\{ E_{ij}\right\} $ of $\Sigma
_{\mathscr{D}}(M)$: The $E_{ij}$'s are the orbits of $\rho _{0}$, $i\in U(%
\mathbb{Z}
_{p^{s}})$ is the determinant of every matrix in $E_{ij}$ and $j$ stands for
an orbit of $\rho _{1}$ (or equivalently of $\rho $). The number of
different values that $j$ can assume is%
\begin{equation}
n_{\rho }=\frac{\left\vert \Sigma _{\mathscr{D}}(M)\right\vert }{\left\vert
U(%
\mathbb{Z}
_{p^{s}})\right\vert ^{2}}.
\end{equation}%
If $2k<s$, then $n_{\rho }=p^{s+2k}$ according to (\ref{nD_k 1}). But if $%
k=s/2$, then%
\begin{equation}
n_{\rho }=\frac{\left\vert \func{GL}(2,%
\mathbb{Z}
_{p^{s}})\right\vert }{\left\vert U(%
\mathbb{Z}
_{p^{s}})\right\vert ^{2}}=\frac{(p^{2s}-p^{2(s-1)})\cdot
(p^{s}-p^{s-1})\cdot p^{s}}{(p^{s}-p^{s-1})^{2}}=p^{2s}+p^{2s-1}>p^{2s}.
\end{equation}

Let $P\in E_{i_{1}j_{1}}$ and $Q\in E_{i_{2}j_{2}}$. On the one hand, $\det
P=i_{1}$ and $\det Q=i_{2}$. On the other hand, $j_{1}=j_{2}$ iff $Q_{1}$
and $Q_{2}$ are proportional to $P_{1}$ and $P_{2}$ respectively. In
passing, we find again that the number of matrices in $\Sigma _{\mathscr{D}%
}(M)$ with some determinant $u$ is the same as the number of matrices in $%
\Sigma _{\mathscr{D}}(M)$ with any other (invertible) determinant $v$.

\section*{Acknowledgments}

The author wishes to thank Subhash Chaturvedi for calling his attention to
all the features about isotropic lines addressed in this communication and
to their involvement in the setting-up of discrete Wigner distributions.

This work is part of the Ph.D. thesis by the author. He is indebted to his
advisor for useful comments and also to Michel Planat and Metod Saniga for
introducing him to finite geometry in the frame of quantum physics.

\bibliographystyle{unsrt}
\bibliography{IsoLines}

\end{document}